\documentclass[twocolumn,showpacs,amsmath,amssymb,prl]{revtex4}
\usepackage{dcolumn}
\usepackage{bm}
\usepackage{psfig}

\newcommand{\BR}{\mathcal{B}}

\newcommand{\gev}{\,\mbox{GeV}}

\newcommand{\ra}{\rightarrow}

\newcommand{\psp}{\psi^{\prime}}
\newcommand{\pspto}{\psi^{\prime}\to}
\newcommand{\jpsi}{J/\psi}
\newcommand{\jpsito}{J/\psi\to}

\newcommand{\psppto}{\psi^{\prime\prime}\to}
\newcommand{\EE}{e^+e^-}
\newcommand{\EETO}{e^+e^-\to}
\newcommand{\pip}{\pi^+}
\newcommand{\pim}{\pi^-}
\newcommand{\piz}{\pi^0}
\newcommand{\threepi}{\pi^+\pi^-\pi^0}

\newcommand{\rhoof}{\rho(1450)}
\newcommand{\rhoos}{\rho(1700)}
\newcommand{\rhoto}{\rho(2150)}

\newcommand{\rhopi}{\rho\pi}
\newcommand{\rholo}{\rho(770)}
\newcommand{\rholopi}{\rholo\pi}

\newcommand{\rhotopi}{\rhoto\pi}

\newcommand{\bfg}{\begin{figure}}
\newcommand{\efg}{\end{figure}}
\newcommand{\bitm}{\begin{itemize}}
\newcommand{\eitm}{\end{itemize}}
\newcommand{\bnum}{\begin{enumerate}}
\newcommand{\enum}{\end{enumerate}}
\newcommand{\btbl}{\begin{table}}
\newcommand{\etbl}{\end{table}}
\newcommand{\btbu}{\begin{tabular}}
\newcommand{\etbu}{\end{tabular}}
\newcommand{\beqns}{\begin{eqnarray*}}
\newcommand{\eeqns}{\end{eqnarray*}}
\begin{document}

\preprint{Draft-PRL}

\title{\boldmath Observation of $\psp$ decays to $\rho(770)\pi$ and
$\rho(2150)\pi$}
\author{ M.~Ablikim$^{1}$, J.~Z.~Bai$^{1}$,
Y.~Ban$^{10}$, J.~G.~Bian$^{1}$, X.~Cai$^{1}$, J.~F.~Chang$^{1}$,
H.~F.~Chen$^{16}$, H.~S.~Chen$^{1}$, H.~X.~Chen$^{1}$,
J.~C.~Chen$^{1}$
% ...
\\(BES Collaboration)\\ } \affiliation{ $^1$
Institute of High Energy Physics, Beijing 100039, People's Republic of
China\\
% ...
$^{10}$ Peking University, Beijing 100871,
People's Republic of China\\
% ...
$^{16}$ University of Science and Technology of China, Hefei 230026,
People's Republic of China\\
% ...
}

\date{\today}

\begin{abstract}

$\pspto \rholopi$ is observed for the first time in a data sample
of 14 million $\psp$ decays collected by the BESII detector at the
BEPC. The branching fraction is measured to be
\(\BR(\pspto\rholopi)=(5.1\pm 0.7\pm 0.8) \times 10^{-5}\), where
the first error is statistical and the second one is systematic. A
high mass excited $\rho$ state with mass around 2.15~GeV/$c^2$ is
also observed with $\BR(\pspto \rho(2150)\pi \to \pi^+ \pi^-
\pi^0) = (19.4 \pm 2.5 ^{+11.2}_{-2.1}) \times 10^{-5}$. The
branching fraction of $\pspto \pi^+ \pi^- \pi^0$ is measured with
improved precision, $\BR(\pspto \pi^+ \pi^- \pi^0) = (18.1 \pm 1.8
\pm 1.9) \times 10^{-5}$. The results may shed light on the
understanding of the longstanding ``$\rhopi$ puzzle" between
$\jpsi$ and $\psp$ hadronic decays.
\end{abstract}

\pacs{13.25.Gv, 12.38.Qk, 14.40.Gx}

\maketitle



From perturbative QCD (pQCD), it is expected that both $\jpsi$ and
$\psp$ decaying into light hadrons are dominated by the
annihilation of $c\bar{c}$ into three gluons or one virtual
photon, with a width proportional to the square of the wave
function at the origin~\cite{appelquist}. This yields the pQCD
``12\% rule'', that is \[ Q_h =\frac{{\cal B}_{\pspto h}}{{\cal
B}_{\jpsito h}} =\frac{{\cal B}_{\pspto \EE}}{{\cal B}_{\jpsito
\EE}} \approx 12\%. \] A large violation of this rule was first
observed in decays to $\rhopi$ and $K^{*+}K^-+c.c.$ by Mark
II~\cite{mk2}, {\it the so called $\rhopi$ puzzle}. Since then BES
has measured many two-body decay modes of the $\psp$; some decays
obey the rule while others violate it~\cite{besres}. There have
been many theoretical efforts trying to solve the
puzzle~\cite{puzzletheory}.  However, none has been accepted as
the solution to the problem.

In the study of the $\rhopi$ puzzle, $\pspto \rhopi$ is one of the
key decay modes and is of great interest to both theorists and
experimentalists. A recent calculation of the $\pspto \rhopi$
branching fraction, done in the framework of SU(3) symmetry, takes
into consideration interference between $\psp$ resonance decays
and the continuum amplitude~\cite{wymphase}; a branching fraction
of $\pspto \rhopi$ around $1\times 10^{-4}$ is predicted with a
large error due to the limited precision for $\psp$ decays into
other vector pseudoscalar (VP) modes. The measurement of the
$\pspto \rhopi$ mode is a direct test of many models proposed to
solve the $\rhopi$ puzzle~\cite{puzzletheory,wymphase}.


The data used for this analysis are taken with the Beijing
Spectrometer (BESII) detector at the Beijing Electron Positron
Collider (BEPC) storage ring operating at the $\psp$ energy. The
number of $\psp$ events is $14 \pm 0.6 $ million~\cite{moxh},
determined from the number of inclusive hadrons, and the luminosity is
$(19.72\pm 0.86)$~pb$^{-1}$ as measured by large angle Bhabha events.

BESII is a conventional solenoidal magnet detector that is
described in detail in Refs.~\cite{bes,bes2}. A 12-layer vertex
chamber (VC) surrounding the beam pipe provides trigger
information. A forty-layer main drift chamber (MDC), located
radially outside the VC, provides trajectory and energy loss
($dE/dx$) information for charged tracks over $85\%$ of the total
solid angle.  The momentum resolution is $\sigma _p/p = 0.017
\sqrt{1+p^2}$ ($p$ in $\hbox{\rm GeV}/c$), and the $dE/dx$
resolution for hadron tracks is $\sim 8\%$. An array of 48
scintillation counters surrounding the MDC  measures the
time-of-flight (TOF) of charged tracks with a resolution of $\sim
200$ ps for hadrons.  Radially outside the TOF system is a 12
radiation length, lead-gas barrel shower counter (BSC).  This
measures the energies of electrons and photons over $\sim 80\%$ of
the total solid angle with an energy resolution of
$\sigma_E/E=22\%/\sqrt{E}$ ($E$ in GeV).  Outside of the
solenoidal coil, which provides a 0.4~Tesla magnetic field over
the tracking volume, is an iron flux return that is instrumented
with three double layers of  counters that identify muons of
momentum greater than 0.5~GeV/$c$.


A phase space Monte Carlo sample of 2 million $\pspto \threepi$ events
is generated for the efficiency determination in the partial wave
analysis (PWA).  Monte Carlo samples of Bhabha, dimuon, and inclusive
hadronic events generated with Lundcharm~\cite{lundcharm} are used for
background studies. The simulation of the detector uses a
Geant3~\cite{geant} based program, which simulates the detector
response, including the interactions of secondary particles with the
detector material. Reasonable agreement between data and Monte Carlo
simulation has been observed in various channels tested, including
$\EETO(\gamma) e^+e^-$, $\EETO (\gamma)\mu^+\mu^-$, $\jpsi\ra
p\bar{p}$, and $\psp \ra \jpsi \pip\pim, \, \jpsi \ra \ell^+\ell^-$
$(\ell=e,\mu)$.


The final state of interest includes two charged pions and one
neutral pion which is reconstructed from two photons. The
candidate events must satisfy the following selection criteria:

\bnum
\item A neutral cluster is considered to be a photon candidate when
  the deposited energy in the BSC is greater than 80~MeV, the angle
  between the nearest charged track and the cluster is greater than
  $16^{\circ}$, the first hit of the cluster is in the beginning six
  radiation lengths of the BSC, and the angle between the cluster development
  direction in the BSC and the photon emission direction is less than
  $37^{\circ}$.  The angle between two nearest photons is required to
  be larger than $7^{\circ}$. The number of photon candidates after
  selection is required to be two.
\item There are two charged tracks in the MDC with net charge zero.
  A track should have a good helix fit and satisfy
  $|\cos\theta|<0.80$, where $\theta$ is the polar angle of the track
  in the MDC.
\item For each charged track, the TOF and $dE/dx$ measurements are
  used to calculate $\chi^2$ values and the corresponding confidence
  levels for the hypotheses that the particle is a pion, kaon, or
  proton ($Prob_\pi$, $Prob_K$, $Prob_p$). At least one charged track
  is required to satisfy $Prob_\pi>Prob_K$ and $Prob_\pi>Prob_p$.
  Radiative Bhabha background is removed by requiring the tracks have
  small $dE/dx$ or small energy deposited in the BSC. Dimuon
  background is removed using the hit information in the muon counter.
\item A four-constraint kinematic fit is performed under the
  hypothesis $\pspto\gamma\gamma\pip\pim$, and the confidence level of
  the fit is required to be greater than 1\%. A Four-constraint
  kinematic fit is also performed under the hypothesis of
  $\pspto\gamma\gamma K^+K^-$, and
  $\chi^2_{\gamma\gamma\pi\pi}<\chi^2_{\gamma\gamma KK}$ is required
  to remove $K^+K^-\piz$ events.
\item To remove background produced by $\psp$ cascade
  decays to $J/\psi$ with $J/\psi \to \mu^+ \mu^-$, the invariant
  mass of $\pip\pim$ is required to be less than
  $2.95~\hbox{GeV}/c^2$.  \enum

\begin{figure}[htbp]
\centerline{ \hbox{\psfig{file=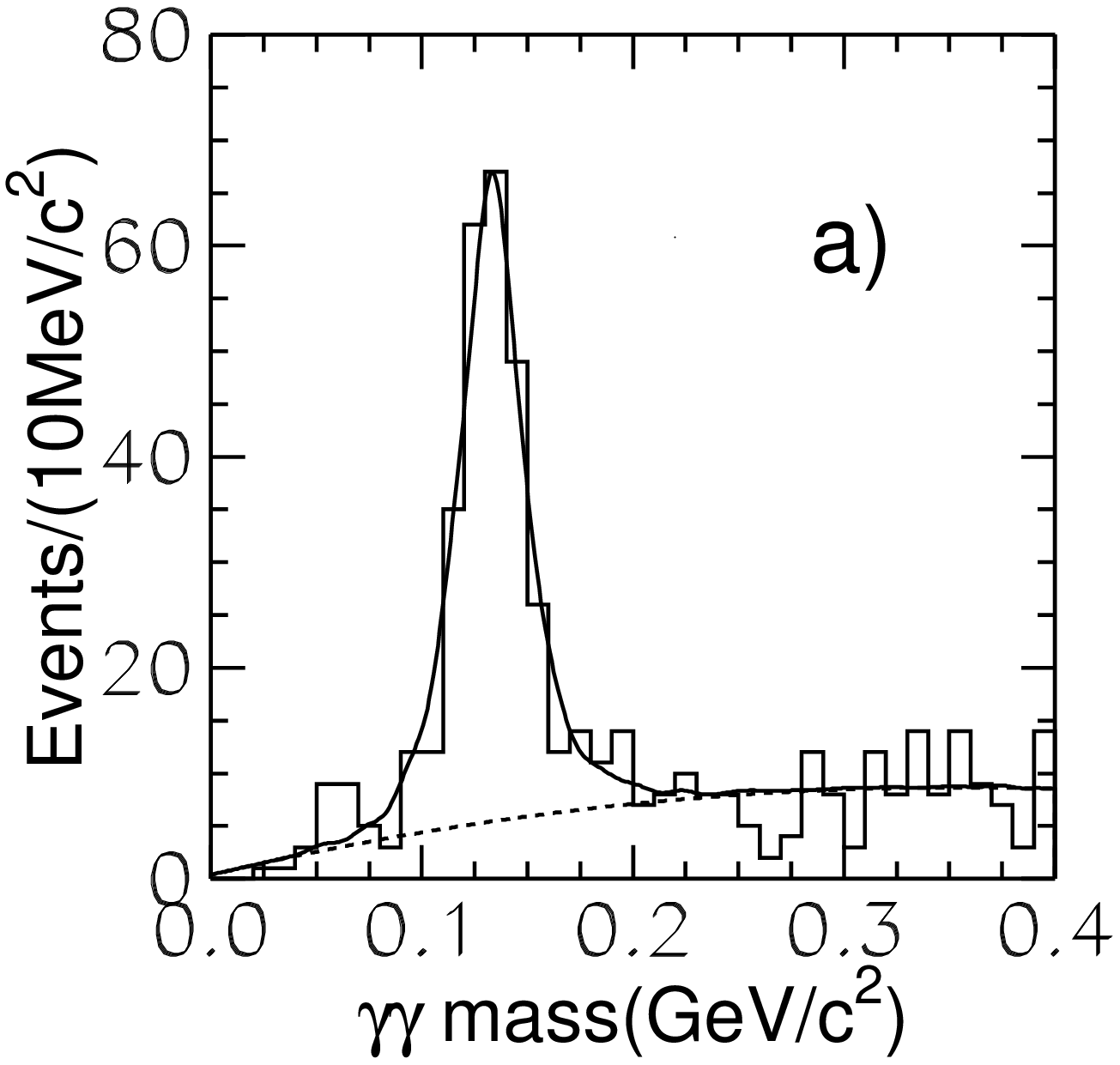,width=4.0cm}}
  \hbox{\psfig{file=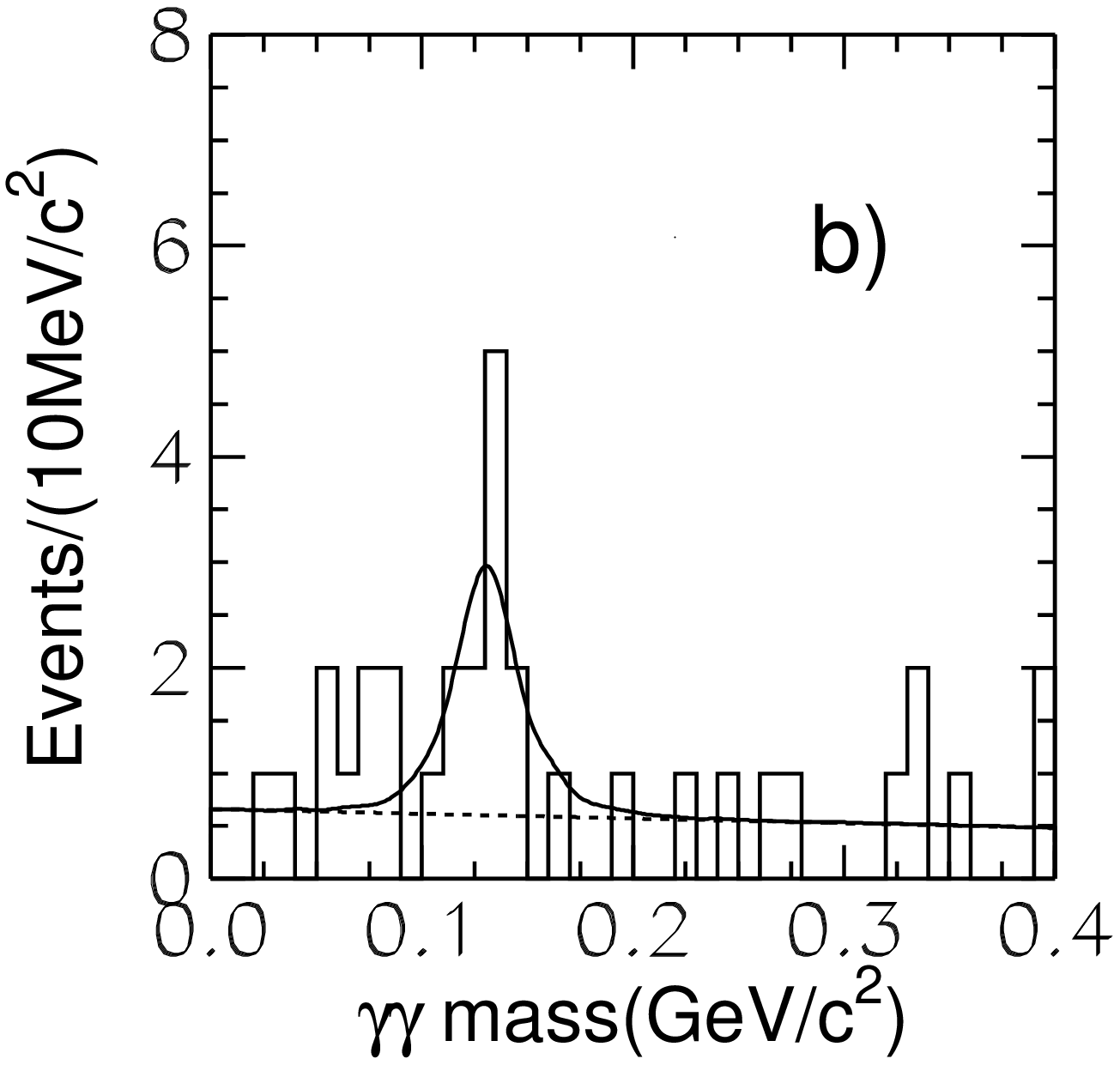,width=4.0cm}}}
\caption{Two photon invariant mass distribution after
final selection for (a) $\psp$ data and (b) continuum data. The
histograms are data, and the curves show the best fits.} \label{mggfit}
\end{figure}

  After applying the above selection criteria, the invariant mass
  distribution of the two photons is shown in Figure~\ref{mggfit}a. A
  clear $\piz$ signal can be seen. A fit to the mass spectrum (shown
  in Figure~\ref{mggfit}a) using a $\piz$ signal shape determined from
  Monte Carlo simulation and a polynomial background yields $260\pm
  19$ $\piz$s.

The contribution from the continuum~\cite{wymphase,wymz} is measured
using $(6.42\pm 0.24)$~pb$^{-1}$ of data taken at $\sqrt{s}=3.65$~GeV
(continuum data). Figure~\ref{mggfit}b shows the $\gamma\gamma$
invariant mass distribution and the fit. The number of $\pi^0$ from
the fit ($10.0\pm 4.2$) is subtracted incoherently from the $\psp$
data after normalizing by the two luminosities. This yields $229\pm
23$ observed $\pspto \threepi$ events.

\begin{figure}[htbp]
\centerline{ \hbox{\psfig{file=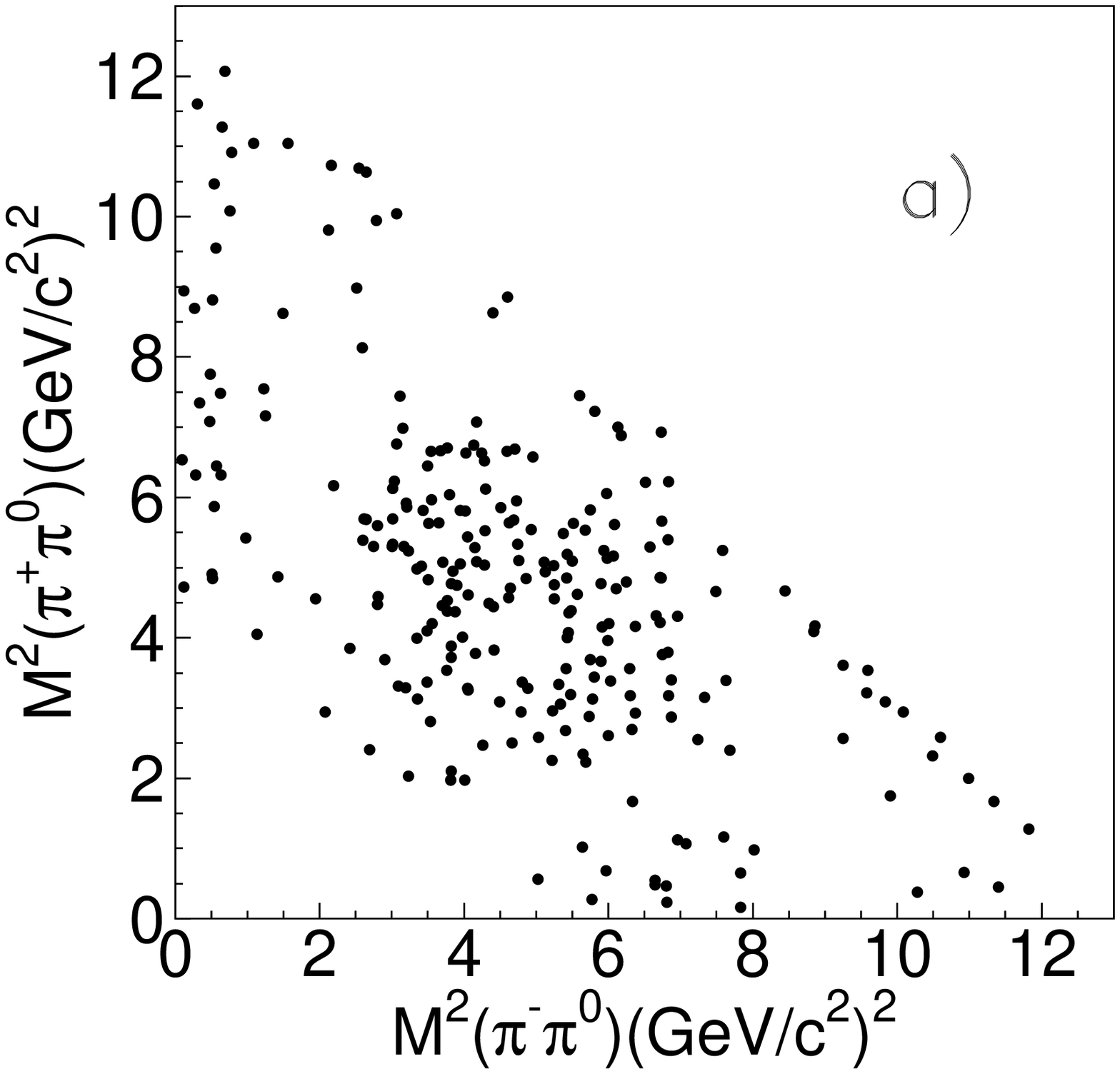,width=4.0cm}}
\hbox{\psfig{file=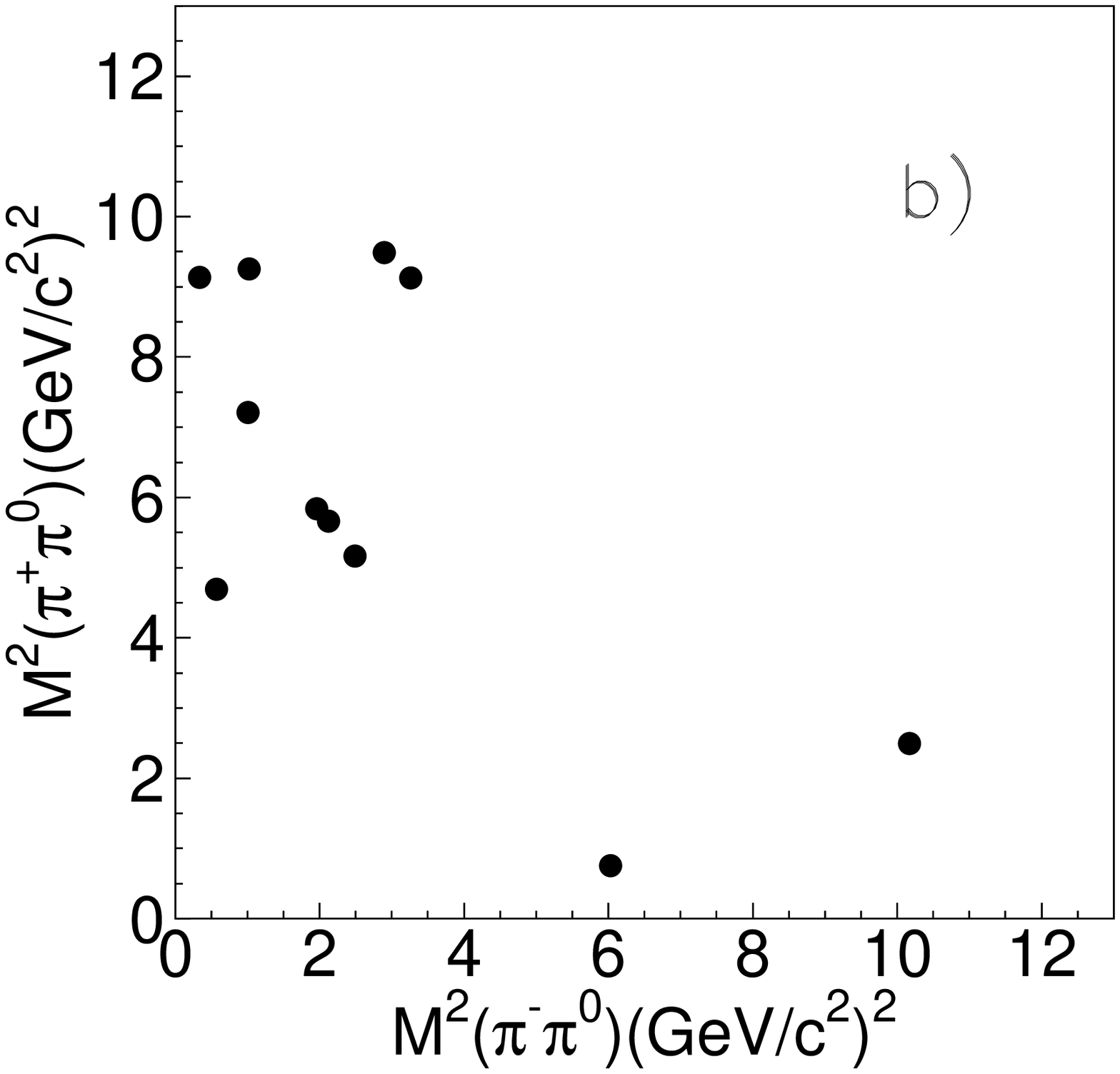,width=4.0cm}}} \caption{Dalitz
plots of $\threepi$ for (a) $\psp$ data and (b) continuum data after
the final selection.} \label{dalitz}
\end{figure}

Dalitz plots of the $\threepi$ system for the $\psp$ and continuum
data are shown in Figure~\ref{dalitz} after requiring the invariant
mass of the two photons lies within $\pm 30$~MeV/$c^2$ of the nominal
$\piz$ mass. (The mass resolution is around 17.5~MeV/$c^2$ from Monte
Carlo simulation.) For the $\psp$ sample, 250 events are obtained with
about 13\% non-$\piz$ background, while for the continuum sample, 11
events are obtained with about 42\% non-$\piz$ background. In $\psp$
decays, besides clear $\rho$ bands at the edges of the Dalitz plot, there
is a cluster of events in the center. This is
very different than the Dalitz plot for $\jpsito \threepi$
decays~\cite{besjpsi3pi}, indicating different decay dynamics between
$\jpsi$ and $\pspto \threepi$. There is no clear intermediate state in the
continuum data.

The selected events are fitted in the helicity amplitude formalism
with an unbinned maximum likelihood method using
MINUIT~\cite{mini}. For the process
\[
  \begin{array}{cccccc}
  \psp(1^-) &  \ra & \rho(1^-)      & +       &  \pi(0^-) &     \\
             &      &\hookrightarrow &\pi(0^-) &  +        &
             \pi(0^-),
  \end{array}
\]
the intensity distribution $dI$ for the final state is written as
\[
dI=\sum_{i=\pm 1}(|A_i|^2+\left|C_i\right|^2)d(LIPS),
\]
where $C_i$ is an incoherent background term, that is assumed to
be either a constant or to have the same angular distribution as
$A_i$. The difference between these two fits is taken as the
systematic error on the background description. $LIPS$ denotes the
Lorentz-invariant phase space, and the amplitude
\[A_i=A^0_i(\pim,\pip)+A^+_i(\pip,\piz)+A^-_i(\piz,\pim), \]
where $i=+1$ or $-1$ is the helicity of the $\psp$, the first pion in
each set of parentheses is the ``designated'' pion, and
 \[A^c_{\pm 1}=B(m^2)\sin\theta_\pi(\cos\phi_\pi\pm i
 \cos\theta\sin\phi_\pi)e^{\pm i\phi}.\]
Here $c=0$, $+1$, or $-1$ is the net charge of the dipion system,
$\theta$ and $\phi$ are the polar and azimuthal angles of the
$\rho$ in the $\psp$ rest frame, $\theta_\pi$ and $\phi_\pi$
are the polar and azimuthal angles of the designated pion in the
$\rho$ rest frame, and $B(m^2)$ describes the dependence of the
amplitude on the dipion mass $m$:
\[
B(m^2)=\frac{BW_{\rho(770)}(m^2)+
       \sum_{j} c_j e^{i\beta_j} BW_j(m^2)}
             {1+\sum_{j} c_j},
\]
where, $BW(m^2)$ is the Breit-Wigner form of the $\rho(770)$ or
its excited states. Here, the Gounaris-Sakurai
parameterization~\cite{gs} is used; $\beta_j$ and $c_j$ are the
relative phase and the relative strength, respectively, between
the excited $\rho$ state $j$ and the $\rho(770)$.

\begin{figure}[htbp]
\centerline{\hbox{ \psfig{file=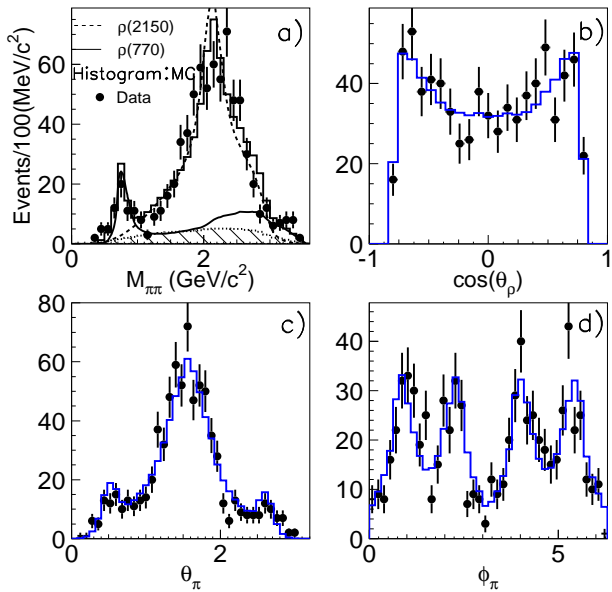,width=8cm}}}
\caption{Comparison between data (dots with error bars) and
the final fit (solid histograms) for
(a) two pion invariant mass, with a solid line for the $\rho(770)\pi$, a
dashed line for the $\rhotopi$,  and a hatched histogram for
background; (b) the $\rho$ polar angle in the $\psp$ rest frame; and
(c) and (d) for the polar and azimuthal angles of the designated $\pi$ in
$\rho$ helicity frame.} \label{fit}
\end{figure}

Since the number of events is limited, the masses and the widths of all states
in the fit are fixed to their PDG values~\cite{pdg}, and the number of
background events is fixed to the number determined from the
$\gamma\gamma$ invariant mass fit. A fit with $\rho(770)$, $\rhoof$,
$\rhoos$ and $\rhoto$ results in insignificant $\rhoof$ and $\rhoos$
contributions. The fit after removing these two components yields a
likelihood decrease of 10.7 with four less free parameters.  The fit
results are shown in Figure~\ref{fit}; the fit describes the data
reasonably well.

The fit parameters and results are given in
Table~\ref{fittab}, where for results without errors, the parameter is
fixed. The fit yields $(28\pm 3)\%$ $\rho(770)\pi$ in all
$\threepi$ events. By comparing the likelihood difference with and
without the $\rho(770)\pi$ in the fit, the significance of
$\rho(770)\pi$ is determined to be $7.8\sigma$. The significance
of $\rhoto\pi$ is larger than $10\sigma$.

\begin{table}
\begin{center}
\caption{\label{fittab}  $\psp\ra\pip\pim\piz$ fitting parameters
and results. For the numbers with no errors, the values are fixed
in the fit. } \btbu {ll} \hline
  Quantity                   &    Fit result \\\hline
 $M_{\rho(770)}(\gev/c^2)$      & $ 0.7711$  \\
 $\Gamma_{\rho(770)}(\gev/c^2)$ & $ 0.1492 $ \\
 $\beta_{\rhoto}(^{\circ})$            &$-102\pm 10$ \\
 $M_{\rhoto}(\gev/c^2)$           & $ 2.149$ \\
 $\Gamma_{\rhoto}(\gev/c^2)$     & $ 0.363$ \\
 $\BR(\threepi)$:$\BR(\rholopi)$:$\BR(\rhoto\pi)$ & 1:$0.28\pm 0.03$:$1.07\pm 0.09$ \\
\hline \etbu
\end{center}
\end{table}

The fit quality is checked using Pearson's $\chi^2$ test by
dividing the Dalitz plots into small areas with at least 20 events
and comparing the number of events between data and normalized
Monte Carlo simulation. A $\chi^2/$ndf $=14.6/7=2.1$ is obtained,
which corresponds to a confidence level of 4\%. A fit with the
$\rhoto$ width free; or a fit with $\rho(770)$, $\rhoof$,
$\rhoos$, and $\rhoto$; or even with an extra excited $\rho$ state
does not improve the fit quality significantly. These cases show
that the change in the number of $\rho(770)\pi$ events is less
than 9.1\%, which is included as part of the systematic error. The
number of $\rhoto\pi$ events increases by 57\% when other excited
$\rho$ states are added in the fit due to the interference; this
is also included in the systematic error.

Using the parameters of the fit in the Monte Carlo
generator, the efficiency of $\pspto \threepi$ is estimated to be
$9.02\%$, and the corresponding
efficiencies for $\rho(770)\pi$ and
$\rhoto\pi$ are $10.54\%$ and $8.70\%$, respectively.


Systematic errors in the $\pspto \threepi$ branching fraction
measurement come from the kinematic fit, the MDC
tracking, charged particle identification, photon identification,
background estimation, continuum subtraction, etc. All
sources considered are listed in Table~\ref{sys}. Most of the
errors are measured using clean exclusive $\jpsi$
and $\psp$ decay samples~\cite{besjpsi3pi,besvt}, while some others
were described above. For the $\rho(770)\pi$ and
$\rhotopi$, the uncertainties of fitting
with different high mass $\rho$ states, etc. are also
included. The total systematic error for $\pspto \threepi$ is
10.5\%, and those for $\pspto \rho(770)\pi$ and $\rhoto\pi$ are
16.0\% and $^{+58.0}_{-10.6}$\%, respectively.

\btbl[htbp]
\begin{center}
\caption{\label{sys}Summary of systematic errors on $\BR(\pspto
\threepi)$.}
 \btbu{l|c}\hline
 Source              & Relative error (\%) \\\hline
Trigger               &0.5\\
MDC tracking          &4.0\\
Kinematic fit        &6.0\\
Photon efficiency         &4.0\\
Number of photons        & 2.0     \\
Background estimation     &3.6\\
Particle ID           &negligible  \\
Total number of $\psp$   &4.0  \\
Continuum subtraction  & 3.0         \\\hline
Total                  & 10.5        \\
\hline \etbu
\end{center}
\etbl


Using the numbers obtained above, the branching fractions of
$\pspto \threepi$, $\rholopi$ and $\rhoto\pi$ are \beqns
\BR(\threepi)=(18.1\pm 1.8\pm 1.9)\times 10^{-5},\\
\BR(\rho(770)\pi\ra\pip\pim\piz)=(5.1\pm 0.7\pm 0.8)\times
10^{-5},\\
\BR(\rhoto\pi\ra\pip\pim\piz)=(19.4\pm 2.5^{+11.2}_{-2.1})\times
10^{-5}, \eeqns where the first errors are statistical and the
second systematic.

Our \(\BR(\pspto\pip\pim\piz)\) agrees with the Mark II~\cite{mk2}
result within $1.8\sigma$, and \(\BR(\pspto\rho(770)\pi) \) is below
the Mark II~\cite{mk2} upper limit and in agreement with one
model prediction~\cite{wymphase}. It should be noted that the
continuum amplitude which is considered incoherently in this
analysis could increase the $\rho(770)\pi$ branching fraction due to
interference with the resonance~\cite{wymphase}.
This should be considered in a higher statistics experiment.

Comparing with the corresponding $\jpsi$ decay branching
fractions, it is found that both $\threepi$ and $\rholopi$ are
highly suppressed compared with the ``12\% rule'', while for
$\rhoto\pi$, there is no measurement in $\jpsi$ decays. It could
be enhanced in $\psp$ decays since the phase space in $\jpsi$
decays is limited due to the large mass of the excited $\rho$
state. It should be noted that using the $\jpsi$ and $\pspto
\rho\pi$ branching fractions, the $\psppto \rho\pi$ branching fraction
and the $\EETO \rho\pi$ cross section  at
$\sqrt{s}=3.773$~GeV can be predicted in the $S$- and $D$-wave mixing
model~\cite{wympspp}, which is proposed as a solution of the $\rho\pi$
puzzle in $\psp$ decays.


In summary, $\pspto \rholopi$ is observed in $\psp$ decays for the
first time, and the branching fraction is measured to be
\(\BR(\pspto\rholopi)=(5.1\pm 0.7\pm 0.8) \times 10^{-5}\). A high
mass excited $\rho$ state at mass around 2.15~GeV/$c^2$ is also
observed with $\BR(\pspto \rho(2150)\pi \to \pi^+ \pi^-  \pi^0) =
(19.4 \pm 2.5 ^{+11.2}_{-2.1}) \times 10^{-5}$.  The results may
shed light on the understanding of the longstanding ``$\rhopi$
puzzle" between $\jpsi$ and $\psp$ hadronic decays.


The BES collaboration thanks the staff of BEPC for their hard
efforts and the members of IHEP computing center for their helpful
assistance. This work is supported in part by the National Natural
Science Foundation of China under contracts Nos. 19991480,
10225524, 10225525, the Chinese Academy of Sciences under contract
No. KJ 95T-03, the 100 Talents Program of CAS under Contract Nos.
U-11, U-24, U-25, and the Knowledge Innovation Project of CAS
under Contract Nos. U-602, U-34 (IHEP); by the National Natural
Science Foundation of China under Contract No. 10175060 (USTC),
and No. 10225522 (Tsinghua University); and by the US Department of
Energy under Contract No. DE-FG03-94ER40833 (U Hawaii).

\end{document}